\documentclass{llncs} %this is for NSS
\usepackage{amssymb}
\usepackage{graphicx}

\usepackage{url}

\usepackage{color}
\usepackage{calc}
\usepackage{siunitx}
\DeclareSIUnit\mt{\milli\tesla} %% A method for say short cut or new unit!
\newcolumntype{P}[1]{>{\centering\arraybackslash}p{#1}}

\usepackage[T1]{fontenc}
\usepackage[ansinew]{inputenc}
\usepackage[english]{babel}
\usepackage{adjustbox}
\usepackage{amsmath,amsfonts,amssymb}
\usepackage{xparse}
\usepackage[section]{placeins} 
\usepackage[misc]{ifsym}
\usepackage{url}

\usepackage[many]{tcolorbox}
\usetikzlibrary{decorations.pathreplacing}

\begin{document}

\mainmatter  % start of an individual contribution

\title{Concealing IMSI in 5G Network Using Identity Based Encryption}
%Concealing IMSI in 5G Network Using Identity Based Cryptography

%\titlerunning{Concealing IMSI Using Identity Based Encryption} 

% their surnames. This ensures that the names appear correctly inlso
\author{Mohsin Khan$^\text{(\Letter)}$%
%%\thanks{Please note that the LNICST Editorial assumes that all authors have used
%%the western naming convention, with given names preceding surnames. This determines
%%the structure of the names in the running heads and the author index.}%
\and Valtteri Niemi\\
}  %

%\authorrunning{Mohsin Khan \and Valtteri Niemi}


% the affiliations are given next
\institute{University of Helsinki, Helsinki, Finland\\
\email{mohsin.khan@helsinki.fi, valtteri.niemi@helsinki.fi}
%P.O. Box 68 (Gustaf H\"allstr\"omin katu 2b)\\
%FI-00014 University of Helsinki\\
%Finland\\
%\url{https://www.cs.helsinki.fi/en}
}

%relationship stu
% corresponding authors can be found in the file "lnicst.dem",
% that is contained in the LNICST LaTeX support package.
%

%%%\toctitle{Lecture Notes in Computer Science}
%%%\tocauthor{Authors' Instructions}
\maketitle

\begin{abstract}
Subscription privacy of a user has been a historical concern with all the previous generation mobile networks, namely, GSM, UMTS, and LTE. While a little improvement have been achieved in securing the privacy of the long-term identity of a subscriber, the so called IMSI catchers are still in existence even in the LTE and advanced LTE networks. Proposals have been published to tackle this problem in 5G based on pseudonyms, and different public-key technologies. This paper looks into the problem of concealing long-term identity of a subscriber and presents a technique based on identity based encryption (IBE) to tackle it. The proposed solution can be extended to a mutual authentication and key agreement protocol between a serving network (SN) and a user equipment (UE). This mutual authentication and key agreement protocol does not need to connect with the home network (HN) on every run. A qualitative comparison of the advantages and disadvantages of different techniques show that our solution is competitive for securing the long-term identity privacy of a user in the 5G network.
\end{abstract}

\section{Introduction}
\label{intro} The NGMN Alliance has pointed out the privacy of a user as a requirement of the 5G network \cite{NGMN_white_paper}. When a user equipment (UE) tries to connect to a network, the UE has to identify itself using an identifier. Once the UE is identified, an authentication protocol is run between the UE and the network. There are two types of attackers against the user privacy. A passive attacker just listens to the radio communication and tries to figure out identity of the user. An active attacker may transmit some radio messages itself. It is easier to protect against a passive attacker than an active attacker. Since 2G (GSM) the network has used temporary identities to protect against passive attackers. However, even in the LTE network the permanent identity is not protected against active attackers.

We discuss solutions to conceal the long-term identifier known as international mobile subscriber identity (IMSI) during the identification phase. These solutions are based on pseudonyms and public-key encryption. The pseudonym based approaches require to maintain a synchronization of pseudonyms between the UE and the HN. We discuss solutions based on certificates and root-key for the category of public key. Public-key based solutions do not require any synchronization. However, the public-key based solutions have higher cost both in terms of communication and computation. 

We propose a novel solution based on identity based encryption (IBE). One additional advantage of our solution is that, it also works as a mutual authentication protocol between SN and UE without the involvement of the HN every time the authentication is needed. This advantage can not be achieved using root-key based approach. This advantage can be achieved using certificate based approach, but it is the heaviest in terms of communication and computation. 
We evaluate our solutions based on the following criteria: (1) Immunity to attackers, (2) Parts of the IMSI concealed, (3 ) Signalling overhead, (4) Latency, (5) PKI complexity, (6) Public-key revocation, etc. The choice of the solution depends on how much we want to achieve. Our solution based on IBE becomes a competitive one by meeting most of the important requirements.

\section{3GPP-defined Aspects of Mobile Networks}
\label{sec:3GPP-defined_aspects_of_mobile_network}
A subscription describes the commercial relationship between the subscriber and the service provider, cf. 3GPP TR 21.905 \cite{TR21905}. A subscription identifier uniquely identifies a subscription in the 3GPP system and is used to access networks based on 3GPP specifications. Subscription privacy refers to the right to protect any information that can be used to identify a subscription to whom such information relates. This definition of privacy suggests to protect any personally identifiable information (PII) from an attacker. While it may be difficult to draw a clear boundary between PII and non-PII, the long-term identifier is surely a PII. 

\subsection{System Overview}
In the case of GSM, 3G (UMTS) and 4G (LTE) networks, IMSI is a long-term identity of a subscriber. An IMSI is usually presented as a $15$ digit number but can be shorter. The first $3$ digits are the mobile country code (MCC), followed by the mobile network code (MNC), either $2$ digits or $3$ digits. The length of the MNC depends on the value of the MCC. The remaining digits are the mobile subscription identification number (MSIN) within the network \cite{TS23003}. 

\begin{figure}
\vspace{-.50 cm}
\begin{center}
  \includegraphics[height= 2.5cm]{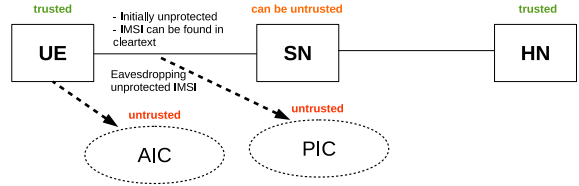}
\caption{High-level security architecture}
\label{fig:security_architecture_abstraction}       % Give a unique label
\end{center}
\end{figure}

In order to present an easily comprehensible discussion, we need to know what are the entities and communication interfaces are involved in this identification process. We also need to know which entities can be entrusted with the IMSI of a subscriber. As the architecture of 5G is yet to be finalized, we present an abstraction of the involved entities and assume that whatever the architecture of 5G will eventually be, it will contain something for each of these entities and something for each of these interfaces. Figure \ref{fig:security_architecture_abstraction} shows the abstraction. The abstraction involves the UE, SN and HN. Note that in a non-roaming situation, the SN and HN are the same network. There are two more entities which are not part of the network but relevant in our discussion, because they attack the network. They are passive IMSI catcher (PIC) and active IMSI cather (AIC). 

The logical interface between UE and SN is initially unprotected. The logical interface between SN and HN is protected. The PICs eavesdrop on the UE-RAN interface when it is unprotected to extract an IMSI. The AICs impersonate a legitimate SN and run a legitimate looking protocol with the UE in order to find out the IMSI.

HN and UE both know the IMSI and they are trusted. Both of PIC and AIC are untrusted. It is in principle possible not to trust SN. However, by other specifications in 3GPP TS 33.106 \cite{TS33106} and TS 33.107 \cite{TS33107}, it is required to reveal IMSI to the SN to enable lawful interception (LI) without involving HN.  
\vspace{-1.50 cm}
\subsection{Current Solution Approach and its Weakness}
One approach of protecting IMSI privacy is to use a temporary identifier instead of the actual IMSI and keep changing the temporary identifier frequently. Note that the temporary identifier has to be assigned confidentially. Different entities of the network may assign different temporary identifiers to the UE. 

In the LTE network, the temporary identifier assigned by an SN is called globally unique temporary identity (GUTI) and the HN does not assign any temporary identifier to the UE. However, during the initial attachment of a UE to the SN, the UE has neither a GUTI nor a security context with the SN that can assign it with a GUTI. Besides, GUTI can be lost by either one or both of the UE and the SN. This would force the UE to reveal its IMSI to the SN to keep itself from permanently locked out of the network.

This problem gives an opportunity to an AIC who impersonates a legitimate SN and forces the UE to run the initial attachment protocol. This also gives an opportunity to a PIC to eavesdrop the IMSI sent in cleartext. Solutions \cite{pseudonym_valtteri_philip,pseudonym_ericsson,CCS15,SSR15} have been proposed by using temporary IMSI known as pseudonym. While these solutions solve the cases of lost and unsynchronised GUTI, they still have the problem of lost or unsynchronised pseudonyms. Public-key technologies have also been considered as potential approach to solve this problem.

\section{Discussion on Different Proposed Solutions}\label{sec:solutions} 
\label{sec:existing_solutions}
Before delving into different proposed those solutions, let us introduce some notation. 
\begin{enumerate}
\item $hnid,snid=MCC||MNC$ identifies the HN and SN respectively
\item $e_A,d_A$ is the public and private key of entity $A$ respectively
\item $\mathcal{X}_{A,B}(e_A,e_B)$ is the certificate of the public key $e_A$ of $A$. The certificate can be verified by anyone who considers $B$ as a root CA using the public key $e_B$. The certificate is a guarantee from B that the public key $e_A$ is owned by $A$ .
\item $E,D$ are encryption and decryption functions so that $D(E(M,K),K) = M$.
\item $S(M,K)$ is the signature of message $M$ signed by the key $K$
\end{enumerate}

\subsection{Solution Based on Pseudonyms:}
\label{sec:pseudonyms}
Pseudonym based solutions have been proposed in \cite{pseudonym_ericsson,pseudonym_valtteri_philip,CCS15,SSR15}. In this kind of solutions, temporary identifiers called pseudonyms are assigned to a UE. Next time when the UE tries to identify itself to an SN, it uses a pseudonym instead of IMSI. Periodically, whenever there is an opportunity, the HN sends a new pseudonym to the UE with confidentiality and integrity protection. One such opportunity could be when the HN sends the authentication vector to an SN.

\subsection{Solution Based on Certificate Based Public-key Cryptography} 
\label{sub_sec:solution_certificate}
Use of certificate based public-key encryption to conceal long-term identity has been suggested in 3GPP TR 33.821 \cite{TR33821}. To use certificate based public-key cryptography, we need to figure out who are the root CAs and who else can be a CA, who own a public key, how a certificate can be revoked, and how the UE can be re-provisioned with a new root certificate if needed. Different solutions can be devised based on the choice of root CAs and other CAs. We provide a high-level description for few variants of certificate based solution.

\subsubsection{Variant 1:}
It uses a global root of trust. There is a global entity trusted by everyone. Using this trusted global entity, a chain of trust can be established. The SN presents the certificate to a UE trying to attach. The UE verifies the certificate. If the verification result is positive, the UE encrypts its IMSI using the public key of the SN and sends to the SN. 

\subsubsection{Variant 2:}
In this variant the HN of a subscriber is the root CA. The HN generates a public-private key pair and generates a certificate of the public key signed by the HN itself. A UE is provisioned with this self signed certificate. An SN interested to serve a UE obtains a certificate $\mathcal{X}_{snid,hnid} (e_{snid},e_{hnid})$. The UE sends $hnid,e_{hnid}$ to the SN. The SN looks up for the certificate $\mathcal{X}_{snid,hnid} (e_{snid},e_{hnid})$. In case it exists at the disposal of the SN, the SN sends it to the UE. The UE verifies the certificate. If the certificate is verified as valid, then the UE sends the IMSI to the SN encrypted by the public key $e_{snid}$ of the SN.

\subsubsection{Variant 3:}
In this variant, there is no other CA than the root CA. Hence the chain of certificates is very short. Only an HN can be a CA. The certificates of all the SNs a UE might visit are pre-provisioned to the UE by the HN. When a UE attempts to attach to an SN, the UE encrypts the IMSI with the public key of the SN which is already provisioned to the UE. If the public key of an SN is revoked, the HN has to provision the revocation to the UE.

\subsection{Solution Based on Root-key based Encryption} 
\label{sub_sec:solution_root-key}
We use only one pair of public-private key pair in this approach. Such a technique has been proposed in 3GPP TR 33.899 in solution \#7.3. This key pair is owned by the HN and we call it to be the root-key. The HN provisions the public key to all its UEs.  Instead of sending the IMSI, the UE encrypts the IMSI with the public root key and sends the result to the SN along with the $hnid$. The SN sends the encrypted IMSI to the HN. The HN decrypts the IMSI and sends the IMSI back to the SN along with an authentication vector (AV).

\subsection{Solution Based on IBE}
In the next section we discuss the basic principles of IBE and present a solution of the identity privacy using IBE. 

\section{Details of the IBE Based Solution} 
\label{sec:solutions_based_on_IBE}
\subsection{How IBE works}
The idea of IBE was proposed by Adi Shamir in 1984 \cite{IBE_shamir}. In IBE, the public and private keys of a receiver are computed from the identity of the receiver in conjunction with the public and private key of a trusted third party respectively. A sender does not need to authenticate the public key of a receiver each time the sender and the receiver agree on a security context. The authenticity of the public key in IBE is guaranteed by the trusted third party. 

Usually in IBE, the trusted third party is known as the private key generator (PKG). The private key of the receiver has to be provisioned to the receiver by the PKG. It is impossible to revoke the public key in IBE unless the identity itself is revoked. Please note that a PKG knows the private keys of all the receivers. As a result a PKG can decrypt any message sent by any sender to any receiver. This implies that there must be a very high level of trust in the PKG.

Dan Boneh and Matthew Franklin published a fully functional IBE scheme in 2003 \cite{IBE_boneh_franklin}. The security of this scheme was based on a natural analogue of the computational Diffie-Hellman assumption. Based on this assumption they showed that the new system has chosen ciphertext security in the random oracle model. To make the revocation of public keys easier, this scheme also suggests to use an expiry time as part of the identity of a receiver. We use this suggestion in our solution. Clifford Cocks present an implementation in 2001 \cite{IBE_clifford} and show that the security of the implementation is related to the difficulty of solving the quadratic residuosity problem.

\subsection{Existing proposals of Using IBE in 5G Network}
RFC 6508 \cite{RFC6508} presents an algorithm SAKKE for establishment of a secret shared value. Applications of SAKKE may include a date-time component in their identity to ensure that identities and hence the corresponding private-keys are only valid for a fixed period of time. Solution \#7.11 in 3GPP TR 33.899 \cite{TR33899} uses IBE to protect the long-term identity according to RFC 6508. However, the solution does not address the issue of revocation of the identity based public-keys. RFC 6507 \cite{RFC6507} describes a certificate-less signature scheme based on IBE. In this scheme a string called public validation token (PVT) randomly chosen by the PKG is assigned to an identity. Both the public and private key of a receiver are computed using the PVT along with the receiver's identity. So, the public key associated with an identity can be revoked by revoking the PVT. Solution \#2.14 in 3GPP TR 33.899 presents an authentication framework based on the signature scheme of RFC 6507 and the authentication protocol EAP-TLS. This solution uses the PVT to revoke the public key associated with an identity. However, in this solution it is not clear how a UE can check if the public key of an SN has been revoked or not.

\subsection{The Proposed Solution}
Next we present a protocol that serves the purposes of both privacy protected identification of UE and mutual authentication between UE and SN. This mutual authentication does not require a contact with the HN each time the protocol is run between a UE and an SN. In our solution we do not use PVT but instead use an expiry time with pre-agreed format. This expiry time can act as the PVT. If the public key of an identity needs to be revoked, the expiry time along with the identity is added to the revocation list. If the identity requires a new public key, the PKG uses another expiry time to compute the private key of the identity. The newly computed private key is then provisioned to the identity along with the the new expiry time. When the expiry time comes, all the public keys computed using the expiry time are automatically revoked. So, the revocation list does not need to include revocations whose expiry time is in the past.

\begin{figure}
\vspace{-.7cm}
\begin{center}
  \includegraphics[height=7cm]{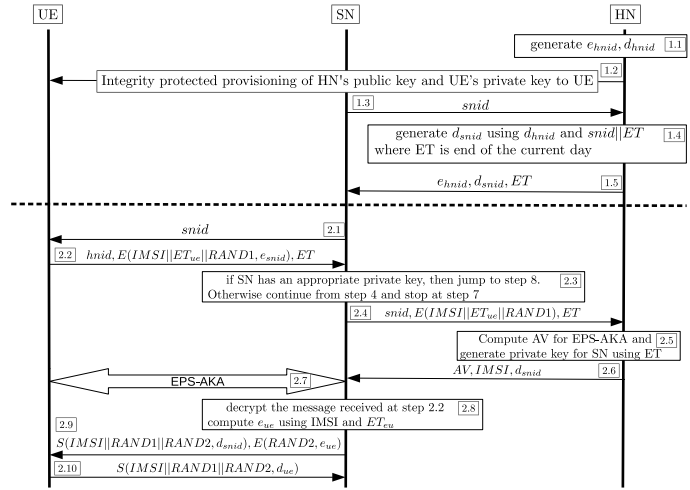}
\vspace{-.5cm}
\caption{Privacy protected UE identification and mutual authentication using IBE}
\label{fig:solution_ibc}       % Give a unique label
\end{center}
\vspace{-1cm}
\end{figure}

\subsubsection{Description of the proposed solution}
The UE's HN acts as the PKG. The solution is pictorially presented in Figure \ref{fig:solution_ibc}. It has two different phases. In the first phase, the key generations and provisioning take place. In the second phase the identification and authentication happens. The description follows:

\begin{itemize}
\item In step $1.1$ the HN generates a public-private key pair $e_{hnid},d_{hnid}$.
\item In $1.2$, the HN provisions the UE with $e_{hnid}$ and $d_{ue}$. $d_{ue}$ is generated using the private key $d_{hnid}$, $IMSI$, and a chosen expiry time $ET_{eu}$. 
\item In  $1.3$, the SN sends the $snid$ to the HN. In  $1.4$ the HN chooses an expiry time $ET$ and $d_{snid}$ is computed considering $snid||ET$ as the SN's identity. 
\item In  $1.5$, the HN sends $d_{snid},ET,e_{hnid}$ to the SN. The SN stores these information in its key-table. 
\item In  $2.1$, the SN broadcasts the $snid$. 
\item In  $2.2$, the UE sends $hnid,E(IMSI||ET_{ue}||RAND1,e_{snid}),ET$ to the SN. 
\item In  $2.3$, the SN looks for a suitable $d_{snid}$ and if found, it jumps to step $2.8$, Otherwise continues from step $2.4$ and stops at $2.7$
\item In $2.4$, SN sends $snid,E(IMSI||ET_{ue}||RAND1),ET$ to HN. 
\item In $2.5$, HN computes the key $d_{snid}$ using $d_{hnid},snid$ and $ET$. Then HN decrypts the IMSI using $d_{snid}$ and  prepares an $AV$.
\item In $2.6$, HN sends  $AV,IMSI,d_{snid}$ to SN. The SN stores $d_{snid},ET$ and in $2.7$ uses the $AV$ to run the EPS-AKA.
\item In $2.8$, SN decrypts the received message and compute $e_{ue}$ using. 
\item In $2.9$ HN sends the signature $S(IMSI||RAND1||RAND2,d_{snid})$ along with $E(RAND2,e_{ue})$ to the UE. The signature is verifiable by $e_{snid}$ in the UE. 
\item In $2.10$, the UE sends the signature $S(IMSI||RAND1||RAND2,d_{ue})$ to the SN which is verifiable by $e_{ue}$. If both UE and SN can verify the signatures as valid, the mutual authentication is completed successfully.
\end{itemize}

Note that the UE and the SN have successfully exchanged two randomly chosen values RAND1 and RAND2 with confidentiality protection. A symmetric key can be computed at both UE and SN using these random values and $e_{hnid}$ using a function like key derivation function used in LTE security. There is also an alternative option of using Diffie-Hellman key exchange protocol. 

\subsubsection{Revocation of Public Keys}
The $ET$ used to generate the public key $d_{snid}$ is quite near in the future, e.g., the day end. So, if the public key needs to be revoked, it would automatically be revoked when the expiration time comes. In this way, a compromised SN would be able mount an attack only for a short period of time. However, the SN would need to get new $d_{snid}$ from the HN before the old $d_{snid}$ expires. 

When the public key of a UE is revoked, the IMSI and relevant $ET$ is stored in a revocation list in the HN.  An SN serving UEs of an HN has a copy of the list. The SN also periodically checks with the HN if there is any new revocations. Before computing the public key of the UE in step $2.8$, the SN checks the revocation list. If it is revoked, the SN discards the message received from the UE and the authentication fails. 

All the entries with expiry time older than current date-time can be removed from the revocation list, hence the revocation list will not grow to a very large size. This frequent private key exchange and refreshing the revocation list would create a bit increased traffic between an SN and HN. On the other hand, this increased traffic is not in the air interface but in the back haul network, which apparently is not very critical.

\section{Comparison of Solutions}
\label{sec:evaluation}
In this paper we have discussed two different categories of solutions: pseudonym based and public-key based. Different solutions \cite{pseudonym_valtteri_philip,pseudonym_ericsson,CCS15,SSR15} have been and more could be devised based on pseudonyms. All these solutions would require the UE and the HN to synchronize their pseudonym states between a UE and the HN.

We have categorized the different public-key technologies into three categories: certificate based, root-key based and identity based. None of them require to maintain synchronization of states between a UE and the HN. But these solutions have some downsides. They need comparatively heavier computational resources, and the ciphertexts are longer which affect the latency. All these Solutions require a mechanism of key revocation.

In certificate based solutions there is a need of a global PKI. However, in some variants of certificate based solutions, the effort to manage a PKI can be reduced significantly. Certificate based solutions require an extra round trip between the UE and SN to exchange and verify the certificate. In a variant of a certificate based solution, this extra round trip could be removed at the expense of provisioning the certificate of an SN to a UE before the UE goes roaming to the SN. All the certificate based solutions have the requirement of exchanging certificates and verifying them. This creates signalling and computational overhead which consequently affect the latency. 

The root-key based solution does not require any extra round trips or certificates, hence it has better signalling and computational overhead compared to certificate based. However, it still suffers from the increased latency in a roaming situation because every authentication needs to travel all the way to the HN. This is because no one else except the HN can decrypt the message sent by the UE. The solution creates also computational pressure in the HN. 

We have proposed a novel solution based on IBE that can both accomplish the identification and mutual authentication. The solution does not need to maintain synchronized states between a UE and the HN. The solution does not require a global PKI and does not need certificates. Unlike the root-key based approach, our solution does not need to involve HN each time authentication is needed. The aforementioned argument makes the IBE based solution a potential candidate to solve the problem in question. In Table \ref{table:comparison}, we present a comparison among the different solutions based on different criteria.

Apparently pseudonym based solution is very good in most of the criteria. One downside of pseudonym based approach is, if the pseudonym is unsynchronized between UE and HN, the user has to visit the HN physically and get back to synchronized state by giving the IMSI in a trusted environment. The need of visiting the HN physically might make the pseudonym based solution a little clumsy. Variant 1 of certificate based approach is good in preventing AIC and also conceals $hnid$. But this is bad in many other important criteria because of exchanging and verifying certificates. Considering the concealment of $hnid$ with a bit less priority, the CertV1 is outperformed by both root-key based and IBE based solution. CertV2 and CertV3 can not even conceal $hnid$. So, the extra overhead of using Certv2 and CertV3 is not worth comparing to root-key and IBE. When comparing IBE and root-key, both of them are almost similar except that IBE based solution is extendible to a mutual authentication protocol between UE and HN. However, Cert1V can also be extended to a mutual authentication protocol. 

\begin{table}
\begin{center}
\caption{Comparative evaluation of the solutions}
\begin{tabular}{ |p{3.5cm}|P{1.25cm}|P{1.25cm}|P{1.25cm}|P{1.25cm}|P{1.5cm}| P{1cm} | }
\hline
\textbf{Criteria} & \textbf{Pseudo} & \textbf{CertV1} & \textbf{CertV2} & \textbf{CertV3} & \textbf{Root-key} & \textbf{IBE}\\
\hline \hline
Immunity to AIC & + - & + & + & + & + & + \\ \hline
Concealing $hnid$ & - & + & - & - & - & - \\ \hline
Signalling overhead & + + & - & - & - & + & + \\ \hline
Computational overhead & + & - & - & - & + & + \\ \hline
Latency while roaming & - & - & -  & - & - & + \\ \hline
Latency while at home & + + & - & -  & - & + & + \\ \hline
PKI effort & + + &  - & + & + & + & + \\ \hline
Key revocation & ++ & - & - & - & + & - \\ \hline
Provisioning effort & + & + & + & - & + & + \\ \hline
Using existing gear & + & - & + & + & + & + \\ \hline
Maturity  & - & + & + & - & + & - \\ \hline
Mutual Authentication & - & + & + & + & - & + \\ \hline
\end{tabular}
\label{table:comparison}
\end{center}
\vspace{-.8cm}
\end{table}
If concealing $hnid$ is essential, then the only applicable solution is Certv1, the certificate based solution with global root of trust. If concealment of $hnid$ can be compromised, then the choice  of the solution depends on the requirement of mutual authentication. If mutual authentication of UE and SN without involving HN is considered important and useful then IBE based solution is the winner. Otherwise, root-key based solution is just enough.

\section{Conclusion}
\label{sec:conclusion}In this paper we have discussed different known approaches to conceal the IMSI. The solutions  are based on pseudonyms and public-key encryption. We have proposed a novel solution based on identity based encryption that serves the purposes of both identification and mutual authentication. We have used expiry time as part of the identity of the entities in the system. We have presented a qualitative comparison between different  solutions. We argue that identity based encryption is a competitive solution when concealing the home network identity is not necessary and mutual authentication in between a user equipment and a serving network is useful without connecting with the home network. The comparison is based on qualitative analysis based on known facts of public-key cryptography.

\subsubsection{Acknowledgement.}
\label{sec:acknowledgement}
We would like to thank Kimmo J\"arvinen for the useful comments and Jarno Niklas Alanko for his valuable feedback.

%\bibliographystyle{splncs}
%\bibliography{mybib}{}

\end{document}